\pdfoutput=1

\documentclass[11pt]{article}

\usepackage[preprint]{acl}
\usepackage{lineno}
\nolinenumbers
\usepackage{times}
\usepackage{latexsym}

\usepackage[T1]{fontenc}

\usepackage[T5]{fontenc}

\usepackage[utf8]{inputenc}

\usepackage{microtype}

\usepackage{inconsolata}

\usepackage{graphicx}
\usepackage{amsmath}

\usepackage{booktabs} 
\usepackage{tabularx} 
\usepackage{hyperref}       
\usepackage{url}            
\usepackage{xurl} 
\usepackage{amsfonts}       
\usepackage{nicefrac}       
\usepackage{microtype}      
\usepackage{xcolor}         
\usepackage{graphicx}       
\usepackage{subfig} 
\usepackage{lscape}
\usepackage{changepage} 
\usepackage{float}  %
\usepackage{multirow}
\usepackage{booktabs}
\usepackage{graphicx}  
\usepackage{titlesec}
\titlespacing*{\section}{0pt}{6pt}{4pt}
\titlespacing*{\subsection}{0pt}{4pt}{2pt}
\titlespacing*{\paragraph}{0pt}{2pt}{0.5em}

\usepackage{enumitem}
\setlist{nosep, topsep=2pt}

\usepackage{placeins}
\title{FPMoE: A Sparse Mixture-of-Experts Approach to Functional Code Generation}

\author{Loc Pham$^{1}$, Nguyet-Anh H. Lang$^{2}$, Thanh Le-Cong$^{3}$, \\
$^{1}$ GreenNode AI, Singapore\\
$^{2}$ Hanoi University of Science and Technology\\
$^{3}$Singapore University of Technology and Design\\
}

\begin{document}
\maketitle
\begin{abstract}

Despite rapid progress in LLM-based code generation, existing models are predominantly trained on imperative languages, leaving functional programming languages (FPLs) such as Haskell, OCaml, and Scala chronically underexplored, with even frontier models performing substantially worse on FPLs. Fine-tuning is a natural remedy, but our experiments show that per-language fine-tuning fails to capture shared functional abstractions, while merged multi-language fine-tuning introduces cross-language interference. To address this, we introduce FPMoE, a lightweight, open-source code generation model built on a sparse Mixture-of-Experts (MoE) architecture with three language-specific routed experts (one each for Haskell, OCaml, and Scala) and a shared expert that captures cross-language functional patterns such as monadic reasoning and type-directed programming. This design resolves both failure modes simultaneously: dedicated experts eliminate interference, while the shared expert preserves abstractions that per-language models miss. On FPEval, FPMoE substantially outperforms fine-tuned baselines and, with only 3B active parameters, matches the performance of much larger models including DeepSeek-Coder-6.7B, Qwen2.5-Coder-14B-Instruct, and Qwen3-Coder-30B-A3B.


\end{abstract}

\section{Introduction}~\label{sec:introduction}
Functional programming is an emerging declarative programming paradigm that conceptualizes computation as the evaluation of mathematical functions. This approach yields software that is modular, deterministic, and amenable to formal reasoning, and has been shown to reduce common defect classes such as race conditions and unintended side effects \cite{Achten2013WhyFP, Hughes1989WhyFP, Ray2014ALS}. These properties have driven the adoption of functional programming languages, e.g., Haskell, OCaml, and Scala , in safety-critical domains such as financial systems \cite{Dijkstra2024FunctionalPI, minsky2011ocaml} and large-scale distributed infrastructure \cite{Zaharia2010SparkCC}.

Despite these benefits, functional programming presents a steep learning curve. Developers accustomed to the imperative style of mainstream languages such as Python and Java must fundamentally shift their mindset to embrace abstractions including recursion, higher-order functions, and monads, which many practitioners find difficult to internalize in practice~\cite{Chakravarty2004TheRA, Tirronen2015UnderstandingBM}

The recent proliferation of LLM-based coding assistants, such as GitHub Copilot \cite{chen2021codex}, Cursor \cite{cursor2023}, and Claude \cite{anthropic2025claude}, offers new opportunities to lower these barriers by supporting developers with tasks such as code generation~\cite{2022AlphaCode, chen2021codex} or bug fixing~\cite{10172803, LeCong2026MemoryEfficient}. Should similar benefits prove transferable to the functional programming paradigm, such assistants may play a meaningful role in reducing the barrier to entry for developers, thereby contributing to broader and more widespread adoption of functional programming in real-world software development.

However, current LLMs fall substantially short in functional programming. A recent evaluation~\cite{Lang2026PerishOF} reveals that models achieve significantly lower performance on functional languages such as Haskell and OCaml compared to imperative counterparts like Java.  Moreover, even functionally correct solutions frequently violate idiomatic functional style, defaulting instead to imperative programming patterns \cite{Lang2026PerishOF}. This gap is unsurprising: dominant pre-training corpora are heavily skewed toward imperative languages, leaving functional programming languages chronically underrepresented.

Crucially, the data imbalance is structural rather than temporary. Functional programming has a small, slow-growing developer community, producing orders of magnitude less open-source code than mainstream languages. Even frontier models trained on far larger corpora than ours still fail systematically on FP idiomaticity (Table~\ref{tab:fpviolation-result}), suggesting that scaling pre-training data alone is unlikely to close the gap.

Fine-tuning offers a natural remedy, yet our initial explorations reveal that standard approaches are insufficient. \textit{Per-language} fine-tuning fails to capture functional abstractions that generalize across languages, effectively learning each language in isolation. \textit{Multi-language} fine-tuning, while exposing the model to all three languages jointly, introduces cross-language interference that degrades per-language performance. Together, these failure modes reveal a fundamental tension: a satisfactory solution must simultaneously preserve language-specific specialization and shared cross-language generalization.

To this end, we introduce \textbf{FPMoE}, a lightweight, open-source code generation model built on a sparse Mixture-of-Experts (MoE) architecture tailored for functional programming. FPMoE comprises three language-specific \textit{routed} experts, each dedicated to Haskell, OCaml, or Scala, complemented by a \textit{shared} expert that continuously captures cross-cutting functional programming patterns, such as monadic reasoning and type-directed programming, common across all three languages. 
This design directly resolves both failure modes above: the dedicated experts eliminate cross-language interference, while the always-active shared expert acquires functional abstractions that are difficult for a per-language model to learn in isolation.


On FPEval~\cite{Lang2026PerishOF}, FPMoE substantially outperforms dense fine-tuned baselines and, at only 3B active parameters, matches or surpasses models several times its size (DeepSeek-Coder-6.7B, Qwen2.5-Coder-14B-Instruct, Qwen3-Coder-30B-A3B). These results suggest that targeted architectural specialization offers a more principled path toward FP code generation than brute-force scaling alone.

In summary, this study  makes the following contributions:

\begin{itemize}
\item \textbf{Problem analysis.} We conduct an empirical analysis of fine-tuning strategies for functional programming code generation, demonstrating that both per-language and multi-language fine-tuning exhibit complementary failure modes that standard approaches cannot resolve simultaneously.



\item \textbf{Model.} We propose FPMoE, which introduces language identity as an explicit routing inductive bias in code MoE---a deliberate departure from purely data-driven routing in prior work (Mixtral, DeepSeekMoE). A shared, always-active expert complements the routed experts to encode cross-language functional abstractions.

\item \textbf{Analysis.} Beyond functional correctness, we evaluate generated code on idiomatic-style adherence (FPViolation), and decompose FPMoE's gains through controlled ablations of each architectural component.
\end{itemize}
\section{Methodology}
\label{sec:methodology}

To develop FPMoE, we adopt a three-stage training pipeline that progressively instills language-specific expertise before promoting cross-language generalization:
\begin{itemize}
    \item \textbf{Expert pre-training} (Section~\ref{sec:phase1}): A base small language model is fine-tuned independently on each target language, yielding three language-specific checkpoints. In parallel, the same base is fine-tuned on the combined corpus to provide a shared-expert initialization.
    \item \textbf{MoE assembly} (Section~\ref{sec:phase3}): The three language-specific checkpoints initialize the routed experts of the MoE; the mixed-language checkpoint initializes the shared expert.
    \item \textbf{Joint MoE fine-tuning} (Section~\ref{sec:phase3}): The assembled MoE---experts, router, shared gate, and backbone---is trained jointly on the combined corpus to consolidate cross-language generalization.
\end{itemize}
\noindent The architectural details of the MoE block itself are deferred to Section~\ref{sec:fpmoe}.

\subsection{Data Collection}
\label{subsec:data_collection}

\paragraph{Data crawling and filtering.}
We collect our training dataset from two publicly available corpora: the CodeParrot GitHub crawl~\cite{codeparrot-github-code-2022} and the OCaml Corpus~\cite{geng2024ocaml}. From these, we retain only repositories written in our target functional programming languages: Haskell, Scala, and OCaml. To reduce processing overhead and ensure data quality, we adopt filtering heuristics inspired by DeepSeek-Coder~\cite{guo2024deepseekcoderlargelanguagemodel} and StarCoder~\cite{li2023starcodersourceyou}. We discard files whose average line length exceeds 100 characters or whose maximum line length exceeds 1{,}000 characters, as such files are typically auto-generated, minified, or otherwise unsuitable for training.

\paragraph{Repository Deduplication and Decontamination.}
We perform deduplication at the repository level rather than the file level, as file-level deduplication may remove individual files while retaining the rest of a repository, potentially corrupting its structural integrity. To prevent data contamination, we apply an $n$-gram filtering procedure to exclude potentially-leaked code snippets from the test set. Specifically, if a piece of code contains a sub-string identical to any sub-string in the test data (FPEval~\cite{Lang2026PerishOF}), we exclude it from our training data.

\subsection{Expert Pre-training}
\label{sec:phase1}

Expert pre-training consists of two parallel supervised fine-tuning runs from the same base checkpoint, each targeting a different level of language specialization.

\paragraph{Language-specific fine-tuning.}
We independently fine-tune the base model on each of the three language-specific corpora (Haskell, OCaml, Scala), yielding three specialized checkpoints. Each is trained exclusively on its corresponding language, allowing it to acquire deep syntactic and semantic knowledge of one language without cross-language interference. These three checkpoints later initialize the routed experts $E_1, E_2, E_3$ of the MoE.

\paragraph{Mixed-language fine-tuning.}
In parallel, we fine-tune a separate model instance on the union of the three corpora. This checkpoint serves two roles:
(i) it is our dense mixed-language baseline (Tables~\ref{tab:experiments-result} and~\ref{tab:shared-expert-scores}), and (ii)~it initializes the shared expert $E_{\text{shared}}$, providing it with prior exposure to cross-language functional patterns before joint training begins.

\paragraph{Repository-level sequence construction.}
A key design choice in our training pipeline is the construction of training sequences at the repository level rather than the individual file level. Concretely, all source files within a repository are concatenated into one contiguous sequence before being fed into the model. This preserves intra-repository dependencies and structural coherence---such as module imports, shared type definitions, and cross-file function references---which are critical for understanding real-world functional programming codebases. This is consistent with recent findings in code language model training~\cite{li2023starcodersourceyou}, which show that repository-level context significantly improves a model's ability to reason about code structure and long-range dependencies.

\paragraph{Training objective.}
Following standard practice in code language modeling, all models are trained using the causal language modeling objective, minimizing the cross-entropy loss over next-token predictions across the full concatenated sequence:
\begin{equation}
    \mathcal{L}_{\text{CE}} = -\frac{1}{T} \sum_{t=1}^{T} \log P_\theta(x_t \mid x_1, x_2, \ldots, x_{t-1})
    \label{eq:ce-loss}
\end{equation}
where $T$ is the sequence length, $x_t$ is the $t$-th token, and $P_\theta(x_t \mid x_1, \ldots, x_{t-1})$ is the predicted probability under model parameters $\theta$.

\subsection{Joint MoE Fine-tuning}
\label{sec:phase3}

After language-specific SFT, we assemble the MoE from the three language-specific checkpoints together with the shared-expert initialization described above. Concretely, the FFN sublayer of each transformer block is replaced by the MoE-FFN block described in Section~\ref{subsec:moe-ffn}; the routed experts inherit the language-specific FFN weights, and the shared expert inherits the mixed-language FFN weights. All other transformer parameters (attention, layer norms, embeddings) are shared across experts and inherited from the base checkpoint. The router and shared-expert gate are initialized randomly.

The assembled MoE is then trained jointly on the combined corpus under the causal language modeling objective augmented with the load-balancing term (Section~\ref{subsec:load-balancing}):
\begin{equation}
    \mathcal{L} = \mathcal{L}_{\text{CE}} + \mathcal{L}_{\text{aux}}
    \label{eq:total-loss}
\end{equation}
This phase has two purposes: (i) the router learns to dispatch tokens to the appropriate language expert, and (ii) the shared expert and routed experts jointly adapt so that the shared expert handles cross-language abstractions while the routed experts retain their language-specific specialization.

\section{FPMoE Architecture}
\label{sec:fpmoe}

\subsection{Design Rationale}
\label{subsec:design-rationale}

Functional programming languages, while sharing foundational concepts such as immutability, higher-order functions, algebraic data types, and type inference, diverge significantly in syntax, runtime semantics, and idiomatic patterns. A monolithic dense model trained on a mixture of these corpora risks averaging over these distinctions, producing mediocre outputs across all languages rather than precise, idiomatic code in any one of them---a failure mode we confirm empirically (Table~\ref{tab:experiments-result}, ``Mixed'' baselines).

The MoE framework offers a natural inductive bias for this regime: routing inputs to specialized sub-networks preserves language-specific representations, while an always-active shared expert lets the model still benefit from cross-language functional reasoning. FPMoE makes this inductive bias explicit through three design choices:
\begin{enumerate}
    \item \textbf{Language-aligned routed experts.} The routed experts are initialized from language-specific checkpoints (Section~\ref{sec:phase1}), giving the router a strong prior to specialize along language identity rather than discover specialization from scratch.
    \item \textbf{An always-active shared expert.} A dedicated expert is applied at every position regardless of routing, encoding cross-language abstractions (monadic reasoning, type-directed programming) that none of the routed experts could be expected to learn in isolation.
    \item \textbf{Mixed-corpus shared-expert initialization.} The shared expert is initialized from a mixed-language fine-tuned checkpoint, so it begins joint training already exposed to cross-language patterns.
\end{enumerate}
\noindent We ablate each of these design choices---the language-aligned routing prior, the shared expert, and the mixed-corpus initialization---in Section~\ref{sec:ablation}.

\begin{figure*}[ht!]
  \centering
  \includegraphics[width=0.9\textwidth]{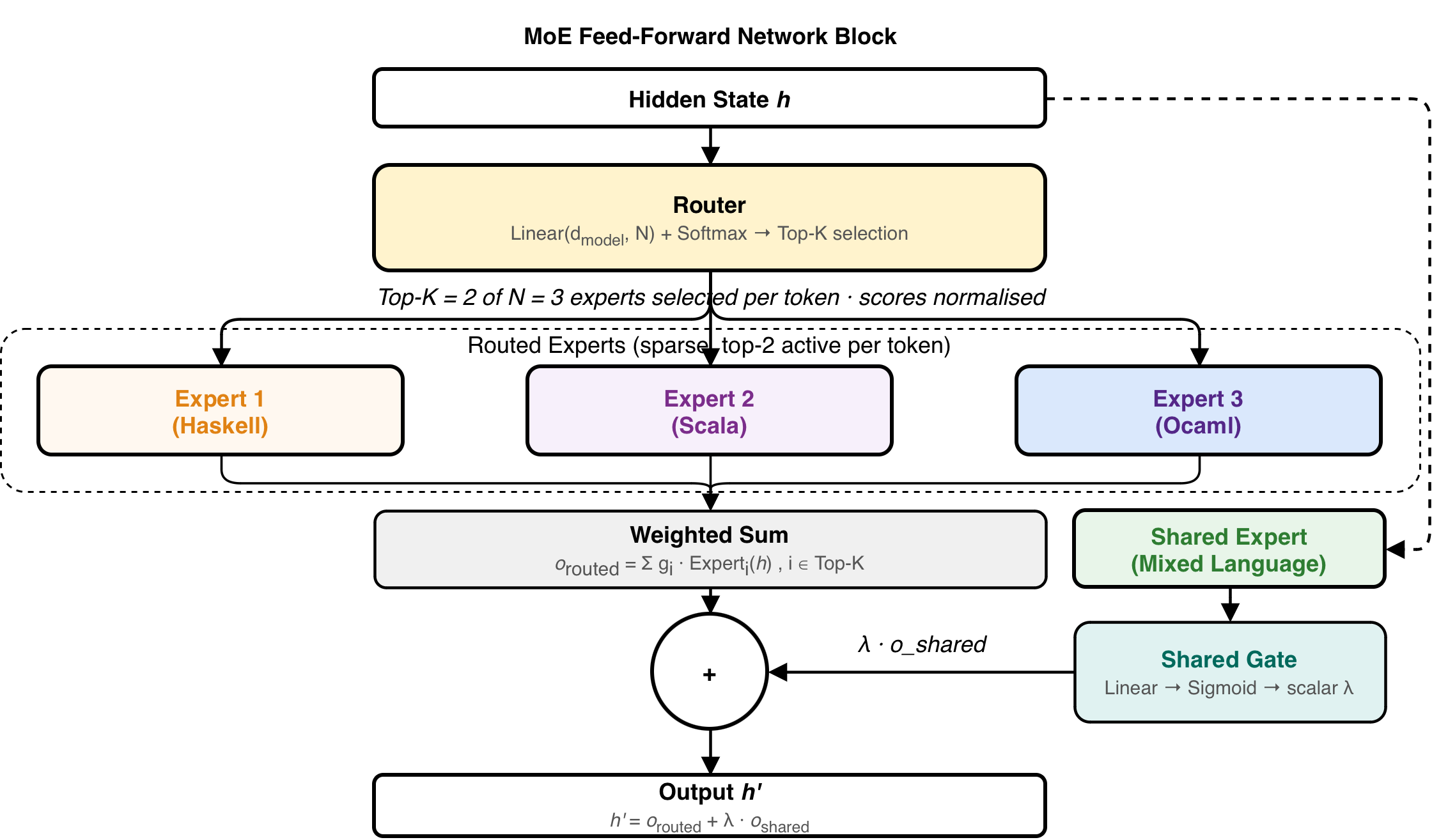}
  \caption{Mixture-of-Experts FFN block. Three language-specific routed experts ($E_1, E_2, E_3$) are dispatched via Top-$K$ routing; the shared expert $E_{\text{shared}}$ is always active and modulated by a learned scalar gate $\lambda$.}
  \label{fig:moe_ffn_block}
\end{figure*}

\subsection{MoE Feed-Forward Block}
\label{subsec:moe-ffn}

Each transformer layer replaces the conventional FFN sublayer with an \textbf{MoE-FFN block} consisting of $N=3$ language-specific routed experts, one always-active shared expert, and two independent gating mechanisms, as illustrated in Figure~\ref{fig:moe_ffn_block}.

\paragraph{Routed experts and router.}
Three expert networks $\{E_1, E_2, E_3\}$ are specialized for Haskell, OCaml, and Scala respectively, each implemented as a two-layer FFN with SwiGLU activation. A lightweight linear router projects the hidden state $h \in \mathbb{R}^{d_{\text{model}}}$ to a distribution over the $N$ routed experts via Top-$K$ selection, producing the routed output:
\begin{align}
    G(h) &= \mathrm{Softmax}(W_g\, h), \label{eq:router} \\
    o_{\text{routed}} &= \sum_{i \,\in\, \text{Top-}K} g_i \cdot E_i(h). \label{eq:routed-output}
\end{align}
In our configurations, $N=3$ and $K=2$.

\paragraph{Shared expert and block output.}
A fourth expert $E_{\text{shared}}$ is always active regardless of routing decisions, encoding cross-language functional patterns such as monadic reasoning and type-directed programming that are consistent across all three languages. Its contribution is modulated by a learned scalar gate $\lambda = \sigma(w_\lambda^\top h + b_\lambda)$, giving the final block output:
\begin{equation}
    h' = o_{\text{routed}} + \lambda \cdot o_{\text{shared}}.
    \label{eq:moe-output}
\end{equation}

\subsection{Model Configurations}
\label{subsec:model-configs}

\begin{table}[t]
\centering
\small
\setlength{\tabcolsep}{4pt}
\begin{tabular}{@{}lcccc@{}}
\toprule
\textbf{Model} & \textbf{Layers} & \textbf{Heads} & \textbf{Experts} & \textbf{Ctx.} \\
\midrule
4x1.5B-A3B & 28 & 12/2 & 4/2 & 32K \\
4x3B-A6B   & 36 & 16/2 & 4/2 & 32K \\
\bottomrule
\end{tabular}
\caption{FPMoE configurations (Qwen2.5-Coder backbone). Both variants use $N=3$ language-specific routed experts plus one always-active shared expert, with Top-$K=2$ routing. ``Heads'' shows Q/KV head counts; ``Experts'' shows Total/Active.}
\label{tab:qwen2.5moe-architecture}
\end{table}

Table~\ref{tab:qwen2.5moe-architecture} summarizes the two FPMoE configurations evaluated in this paper.

\subsection{Load Balancing}
\label{subsec:load-balancing}

To prevent \emph{routing collapse}---where the router converges to dispatching all tokens to a single expert---we apply the auxiliary load-balancing loss of \citet{fedus2022switchtransformersscalingtrillion}:
\begin{equation}
    \mathcal{L}_{\text{aux}} = \alpha \cdot N \sum_{i=1}^{N} f_i \cdot p_i,
    \label{eq:aux-loss}
\end{equation}
where $f_i$ is the fraction of tokens dispatched to routed expert $i$, $p_i$ is the mean router probability for expert $i$, $N=3$ is the number of routed experts, and $\alpha=10^{-2}$ is the load-balancing coefficient. The shared expert is unrouted by construction and not subject to this term.
\section{Experiments}\label{sec:experiments}



\subsection{Experiment Setup}
\textbf{Base Models}. We conduct experiments on two model backbones of different scales: Qwen2.5-Coder-1.5B and Qwen2.5-Coder-3B (see detailed architecture information in Table~\ref{tab:qwen2.5moe-architecture}), yielding FPMoE-Qwen2.5-4x1.5B-A3B and FPMoE-Qwen2.5-4x3B-A6B, respectively. Both share the same MoE configuration of 4 total experts with 2 activated per token.

\noindent\textbf{Training Stages.} Following Section~\ref{sec:methodology}, training proceeds in three steps:
\begin{itemize}
    \item Language-specific SFT: the base model is independently fine-tuned on each of the three language corpora (Haskell, OCaml, Scala), yielding three expert checkpoints.
    \item Mixed-language SFT: a separate model instance is fine-tuned on the combined corpus to serve as the shared expert initialization.
    \item Joint MoE fine-tuning: Three language-specific experts and the shared expert are merged into the MoE architecture, then trained jointly to optimize the router alongside the combined causal LM and load balancing loss Eq~\ref{eq:aux-loss}.
\end{itemize}

\paragraph{Evaluation Benchmark.}
We evaluate on FPEval \cite{Lang2026PerishOF}, a benchmark of 721 programming tasks stratified across three difficulty levels (184 easy, 346 medium, and 191 hard) in Haskell, OCaml, and Scala, with Java as an imperative baseline. Each task is paired with a comprehensive test suite, and evaluation is complemented by language-specific static analysis tools, enabling assessment of both functional correctness and adherence of LLM-generated code to best practices in functional programming.

\begin{table}[!tb]

\centering
\small
\resizebox{0.45\textwidth}{!}{%
\begin{tabular}{p{0.1cm}lccc}
\toprule
\textbf{Cat} & \textbf{Model} & \textbf{Haskell} & \textbf{OCaml} & \textbf{Scala} \\
\midrule

\multirow{4}{*}{\rotatebox[origin=c]{90}{FT 1.5B}}
 & Qwen2.5-Coder-1.5B + Haskell  & 1.39\% & --     & --     \\
 & Qwen2.5-Coder-1.5B + OCaml    & --     & 1.11\% & --     \\
 & Qwen2.5-Coder-1.5B + Scala    & --     & --     & 1.25\% \\
 & Qwen2.5-Coder-1.5B + Mixed      & 0.42\% & 1.11\% & 3.19\%     \\
\midrule

\multirow{4}{*}{\rotatebox[origin=c]{90}{FT 3B}}
 & Qwen2.5-Coder-3B + Haskell    & 3.36\% & --     & --     \\
 & Qwen2.5-Coder-3B + OCaml      & --     & 3.19\% & --     \\
 & Qwen2.5-Coder-3B + Scala      & --     & --     & 11.65\% \\
 & Qwen2.5-Coder-3B + Mixed        & 4.46\%     & 0.97\% & 11.1\%     \\
\midrule
\multirow{2}{*}{\rotatebox[origin=c]{90}{\textbf{MoE FT}}}
 & FPMoE-Qwen2.5-4x1.5B-A3B    & \textbf{7.37\%} & 6.8\%     & \textbf{13.31\%}     \\[6pt]
 & FPMoE-Qwen2.5-4x3B-A6B &    \textbf{11.00\%} & \textbf{12.34\%} & \textbf{21.08\%} \\[6pt]
\midrule

\multirow{9}{*}{\rotatebox[origin=c]{90}{Base}}
 & DeepSeek-Coder-1.3B           & 0.56\% & 0.28\%  & 2.64\%  \\
 & DeepSeek-Coder-6.7B           & 2.92\% & 2.36\%  & 8.46\%  \\
 & Qwen2.5-Coder-1.5B   & 1.11\% & 0.14\% & 3.46\% \\
& Qwen2.5-Coder-3B  & 4.9\% & 2.08\% & 12.62\% \\
 & Qwen2.5-Coder-14B             & 14.50\% & 17.61\% & 30.65\% \\
 & Qwen3-Coder-30B-A3B           & 20.45\% & 25.14\% & 36.73\% \\
 & DeepSeek-V3.2          & 34.45\% & 46.31\% & 44.35\% \\
 & Devstral Small 1.1          & 10.77\% & 9.48\% & 24.97\% \\
 & Qwen3-Coder-Next          & 12.75\% & 31.94\% & 33.33\% \\
 \midrule
\multirow{3}{*}{\rotatebox[origin=c]{90}{Prop Model}}
 & GPT-3.5           & 14.15\% & 9.43\% & 19.28\% \\[1pt]
 & GPT-4o           & 27.18\% & 36.2\% & 38.83\% \\[1pt]
 & GPT-5           & 42.34\% & 52.16\% & 58.36\% \\[1pt]
 & Claude Opus 4.6           & 40.07\% & 73.36\% & 41.36\% \\[1pt]
\bottomrule
\end{tabular}
}
\caption{Pass rate (\%) across functional programming languages. FT = Fine-tuned. Prop = Propriety. \\ MoE FT = Our work}
\label{tab:experiments-result}
\end{table}
\FloatBarrier
\vspace{4em}
\subsection{Main Results}
Table~\ref{tab:experiments-result} reports Pass@1 scores across Haskell, OCaml, and Scala for all model configurations.

\textbf{MoE vs. Dense Fine-tuning.} Both FPMoE variants consistently outperform their dense fine-tuned counterparts at comparable active parameter budgets, with gains of $5$--$11\times$ at 1.5B-active-parameter scale and $3$--$7\times$ at 3B scale (Table~\ref{tab:experiments-result}).

\textbf{Parameter Efficiency.}
As shown in Table~\ref{tab:experiments-result}, FPMoE demonstrates strong parameter
efficiency at both scales. FPMoE-Qwen2.5-4x1.5B-A3B (3B active parameters) surpasses
DeepSeek-Coder-6.7B by margins of $2.5\times$, $2.9\times$, and $1.6\times$ on Haskell,
OCaml, and Scala respectively, despite having less than half its active parameters.
At the larger scale, FPMoE-Qwen2.5-4x3B-A6B approaches Qwen2.5-Coder-14B
(11.00\%, 12.34\%, 21.08\% vs.\ 14.50\%, 17.61\%, 30.65\%) while activating less than
half its parameters, and surpasses GPT-3.5 on both OCaml and Scala. These results
suggest that targeted MoE specialization can substitute effectively for brute-force
scale in functional programming.

\paragraph{Per-language Trends.}
Scala consistently scores highest, likely reflecting its larger training corpus share; Haskell remains the most challenging. OCaml shows the sharpest sensitivity to cross-language interference (mixed 3B baseline degrades $3.3\times$ vs.\ OCaml-specific), which FPMoE's routing successfully mitigates.

\textbf{Architectural specialization vs. data scaling}. A natural alternative to FPMoE is simply pre-training on more functional programming data. We argue the two are complementary, but architectural specialization is the better-suited intervention for FP for three reasons. First, the FP data ecosystem is structurally small: our Haskell and OCaml corpora total approximately 1B tokens combined, while mainstream Python corpora exceed this by orders of magnitude, and the size of the FP developer community~\citep{Chakravarty2004TheRA, Tirronen2015UnderstandingBM} means this gap will not close on its own. Second, frontier models trained on far larger and better-curated corpora than ours still exhibit FPViolation rates of 80–94\% on Haskell and Scala (Table ~\ref{tab:fpviolation-result}); if data scaling alone could close the idiomaticity gap, frontier-scale training would already have done so. Third, our own dense Mixed baseline (Qwen2.5-Coder-3B + Mixed: 4.46\%/0.97\%/11.10\%) is effectively the ``add more FP data'' approach at our scale, and it substantially underperforms FPMoE while still exhibiting cross-language interference, because mixing data does not preserve language-specific abstractions. Targeted architectural specialization uses every available FP token more efficiently than mixed-corpus fine-tuning, which is what makes it the appropriate intervention for data-scarce paradigms like FP.


\paragraph{Comparison with Proprietary Models.}
FPMoE-Qwen2.5-4x3B-A6B surpasses GPT-3.5~\cite{openai2022chatgpt} on OCaml and Scala while remaining slightly behind on Haskell---a notable result for a 6B-active-parameter open-source model. A substantial gap remains relative to frontier models (GPT-5, Claude Opus 4.6), reflecting the orders-of-magnitude difference in scale and proprietary resources; we view this as a future-scaling target rather than a fundamental limitation.


\section{Ablation Study}
\label{sec:ablation}
\subsection{Ablation Configurations}


We isolate the contribution of each FPMoE design element through five controlled ablation configurations (Table~\ref{tab:shared-expert-ablation}), holding all other components fixed across runs.

\begin{table}[h]
\centering
\scriptsize
\caption{Ablation configurations. Configs A--C vary the shared expert initialization
with routed experts held fixed (language-specific warm-up). Configs D and E remove
the shared expert and the language-identity inductive bias, respectively.}
\label{tab:shared-expert-ablation}
\setlength{\tabcolsep}{2pt}
\begin{tabular}{@{}cll@{}}
\toprule
\textbf{Config} & \textbf{Label} & \textbf{Description} \\
\midrule
A & FPMoE (ours)        & Mixed FP corpus pre-training            \\
B & Base + Fine-tuned   & General base model, jointly tuned       \\
C & Base + Frozen       & General base model, frozen              \\
D & No Shared Expert    & Routed experts only (no shared pathway) \\
E & Relaxed MoE         & All experts from same mixed checkpoint  \\
\bottomrule
\end{tabular}
\vspace{-3mm}
\end{table}

Table~\ref{tab:shared-expert-ablation} summarizes the five primary configurations 
evaluated in this ablation.

\paragraph{Configuration Descriptions}
For Configs A–C, the central tension is the trade-off between \textit{ corpus-specific pre-training} and \textit{general reasoning capacity} for the shared expert. Configs D and E additionally test whether the shared expert and the language-identity inductive bias are \textit{individually necessary}.

\paragraph{Config A --- Mixed FP Pre-training.}
The shared expert is pre-trained on a corpus spanning all three FPLs before joint MoE fine-tuning, granting explicit exposure to cross-language patterns (monadic reasoning, algebraic data types).

\paragraph{Config B --- Base Model Fine-tuned.}
The shared expert is initialized from the general-purpose base model and adapts during joint MoE fine-tuning.

\paragraph{Config C --- Base Model Frozen.}
The shared expert is initialized from the base model with parameters frozen throughout fine-tuning.

\paragraph{Config D --- No Shared Expert.}
The shared expert is removed entirely (intermediate size set to zero), retaining only the three routed experts.

\paragraph{Config E --- Relaxed MoE.}
All four experts are initialized identically from the same dense mixed-language checkpoint, with no language-specific warm-up; the router discovers specialization purely from data.

\subsection{Benchmark Results}
Table~\ref{tab:shared-expert-scores} presents the Pass@1 results across all ablation configurations. \textbf{Config A} consistently outperforms every alternative; we decompose the contribution of each design element below.

\paragraph{Shared expert initialization (A vs.\ B, C).}
Config A markedly outperforms both B and C across all three languages. Initializing the shared expert from a general-purpose base model is insufficient even when allowed to adapt during training (Config B); without prior exposure to cross-language FP patterns, the shared expert competes with the routed experts rather than complementing them. Frozen general-purpose priors (Config C) perform worst. Notably, both B and C fall below the dense fine-tuned baselines, indicating that poor shared-expert initialization can actively degrade MoE performance.

\paragraph{Contribution of the shared expert (A vs.\ D).} Removing the shared expert entirely (\textbf{Config D}) costs $1.72$ average Pass@1 points relative to Config A ($-0.90$ Haskell, $-1.53$ OCaml, $-2.73$ Scala), confirming that the shared expert provides measurable signal beyond routed specialization. The gap is modest on correctness but larger on idiomaticity (Section~\ref{subsec:fp-violation}), consistent with our hypothesis that the shared expert primarily encodes cross-language abstractions (monadic reasoning, type-directed patterns) that drive idiomatic style rather than raw test-passing. Notably, Config D still substantially outperforms the dense Mixed baseline ($+7.58$ avg.\ points), indicating that routed experts alone capture most of FPMoE's gain, with the shared expert refining the result.

\paragraph{Contribution of language-aligned routing (A vs.\ E).} Without language-specific warm-up (\textbf{Config E}), the router must discover specialization purely from data and trails Config A by $3.24$ average points ($8.60/8.68/17.35$ vs.\ $11.00/12.34/21.08$). The largest drop is on OCaml ($-3.66$) --- the language with the strongest cross-language interference signal in Table~\ref{tab:experiments-result}. This pattern directly supports our central architectural claim: FPMoE's gains are not attributable to sparse MoE capacity alone. Without a language-aligned signal at initialization, the router recovers only about two-thirds of FPMoE's gain over the dense Mixed baseline ($+6.06$ vs.\ $+9.30$ avg.\ points).

\paragraph{Full decomposition.}
Combining all four ablations yields Dense Mixed (5.51) < E (11.57) < D (13.09) < A (14.81): the language-identity inductive bias and the shared expert contribute additively.

\begin{table*}[ht]
\centering
\caption{Pass@1 (\%) on FPEval across ablation configurations (FPMoE-Qwen2.5-4x3B-A6B backbone). Config A is full FPMoE; B--E each remove or alter one design element. The decomposition Dense Mixed $<$ E $<$ D $<$ A shows that the language-identity inductive bias and the shared expert contribute additively.}
\label{tab:shared-expert-scores}
\resizebox{0.8\textwidth}{!}{
\begin{tabular}{clcccc}
\toprule
\textbf{Config} & \textbf{Label} & \textbf{Haskell} & \textbf{OCaml} & \textbf{Scala} & \textbf{Avg.} \\
\midrule
A & FPMoE (ours) --- Mixed FP Pre-training  & \textbf{11.00\%} & \textbf{12.34\%} & \textbf{21.08\%} & \textbf{14.81\%} \\
B & Base Model + Fine-tuned                 & 1.67\%           & 1.66\%           & 7.21\%           & 3.51\%           \\
C & Base Model + Frozen                     & 0.69\%           & 0.55\%           & 0.83\%           & 0.69\%           \\
D & No Shared Expert                        & 10.10\%          & 10.81\%          & 18.35\%          & 13.09\%          \\
E & Relaxed MoE (no language warm-up)       & 8.60\%           & 8.68\%           & 17.35\%          & 11.57\%          \\
\midrule
\multicolumn{6}{l}{\textit{Dense Fine-tuned baselines}} \\
\midrule
— & Qwen2.5-Coder-3B (Language specific)    & 3.36\%           & 3.19\%           & 11.65\%          & 6.07\%           \\
— & Qwen2.5-Coder-3B (Mixed language)       & 4.46\%           & 0.97\%           & 11.10\%          & 5.51\%           \\
\midrule
\multicolumn{6}{l}{\textit{Large Model References}} \\
\midrule
— & DeepSeek-Coder-6.7B                     & 2.92\%           & 2.36\%           & 8.46\%           & 4.58\%           \\
— & Qwen2.5-Coder-14B                       & 14.50\%          & 17.61\%          & 30.65\%          & 20.92\%          \\
— & Qwen3-Coder-30B-A3B                     & 20.45\%          & 25.14\%          & 36.73\%          & 27.44\%          \\
\bottomrule
\end{tabular}
}
\vspace{-3mm}
\end{table*}

\subsection{Adherence to Functional Programming}
\label{subsec:fp-violation} 
Beyond correctness failures, we identify a class of errors we term \textbf{FPViolation}, cases where LLM-generated code passes test validation but violates core functional programming principles, such as immutability, referential transparency, and idiomatic use of higher-order functions~\citep{Lang2026PerishOF}. Table~\ref{tab:fpviolation-result} reports the Pass@1 rate of code with non-violated patterns (lower is better), revealing a nuanced picture that complements Table~\ref{tab:experiments-result} and Table~\ref{tab:shared-expert-scores}.

\paragraph{FPMoE achieves strong correctness-to-idiomaticity balance.} FPMoE-Qwen2.5-4x3B-A6B records competitive FPViolation rates across all three languages, despite lower raw pass rates than frontier models. Notably, GPT-5 and DeepSeek-V3.2, despite leading in correctness, exhibit among the highest violation rates, suggesting that scale alone does not resolve the tendency to generate imperative-style solutions that merely happen to pass tests.

\paragraph{FPMoE's idiomaticity stems from architectural design.} 
Unlike dense fine-tuned baselines, FPMoE's dedicated routing channels language-specific patterns through specialized experts, actively reducing idiomatic violations. Pass rate alone is therefore insufficient for evaluating FP code generation.

\begin{table}[ht]
\small
\centering
\caption{Proportion of LLM-generated code violating functional programming best practices across Haskell, OCaml, and Scala on FPEval (lower is better).}
\begin{tabular}{l@{\hspace{8pt}}c@{\hspace{8pt}}c@{\hspace{8pt}}c}
\toprule
\textbf{Model} & \textbf{Haskell} & \textbf{OCaml} & \textbf{Scala} \\
\midrule
\multicolumn{4}{l}{\textit{FPMoE (our work)}} \\
\midrule
FPMoE-Qwen2.5-4x1.5B-A3B  & 62.55\% & 42.72\% & 66.16\% \\
FPMoE-Qwen2.5-4x3B-A6B              & 70.18\%           & 57.42\%           & 66.16\%           \\
\midrule
\multicolumn{4}{l}{\textit{Large Model References}} \\
\midrule

Qwen3-Coder-30B-A3B       & 77.25\% & 63.11\% & 81.28\% \\
Qwen3-Coder-Next          & 64.83\% & 56.79\% & 70.41\% \\
DeepSeek-V3.2             & 82.94\% & 64.27\% & 89.73\% \\
GPT-4o                    & 63.18\% & 42.35\% & 87.69\% \\
GPT-3.5                   & 53.42\% & 57.13\% & 93.85\% \\
GPT-5                     & 87.91\% & 79.64\% & 93.72\% \\
Claude Opus 4.6           & 80.27\% & 81.58\% & 78.94\% \\
\bottomrule
\label{tab:fpviolation-result}
\end{tabular}
\vspace{-5mm}
\end{table}



\section{Related Work}

\paragraph{Automation in Functional Programming.}
Several works have investigated the automation of functional programming tasks. \citet{gissurarson2022propr} proposed PropR, a property-based automated program repair framework for Haskell that leverages type-driven synthesis to fix bugs. \citet{van2024investigating} fine-tuned UniXCoder and CodeGPT specifically for Haskell code completion, demonstrating that language-specific fine-tuning can significantly outperform general-purpose LLMs in this domain. Our work differs from theirs in two key aspects. First, we target three functional languages simultaneously---Haskell, OCaml, and Scala---rather than Haskell alone, exposing the additional challenge of cross-language interference that single-language approaches cannot address. Second, their fine-tuned models are not publicly available, precluding direct comparison; we instead treat dense per-language and multi-language fine-tuning as strong and reproducible baselines, which we show are systematically outperformed by FPMoE.
More recently, FPEval~\cite{Lang2026PerishOF} conducted the first holistic evaluation of frontier LLMs on functional programming, revealing a consistent performance gap between functional and imperative languages and identifying pervasive idiomatic-style violations in LLM-generated functional code. Building on these findings, FPMoE introduces a sparse MoE architecture specifically designed to improve LLM code generation across functional programming languages.

\paragraph{Mixture-of-Experts (MoE).}
Starting from \citet{Jacobs1991AdaptiveMO}, MoE models improve capacity by routing inputs to specialized sub-networks via a gating mechanism. \citet{shazeer2017outrageouslylargeneuralnetworks} scaled this idea with the Sparsely-Gated MoE layer, and Switch Transformer~\cite{fedus2022} further simplified routing to a Top-1 scheme while introducing an auxiliary load-balancing loss to prevent routing collapse, a technique we adopt in FPMoE. Subsequent architectures such as Mixtral~\cite{jiang2024mixtral} and DeepSeekMoE~\cite{dai2024deepseekmoe} refined this with Top-$K$ routing and fine-grained expert decomposition, achieving strong performance at reduced active-parameter budgets.

In contrast to prior MoE work in code generation, which relies on purely data-driven routing, FPMoE introduces an inductive bias aligned with the domain: language identity directly informs expert assignment, and a dedicated shared expert---initialized on a cross-language corpus and always active---captures functional programming abstractions common across all three target languages, addressing the failure modes identified in Section~\ref{sec:methodology}.

\section{Conclusion}


We presented FPMoE, a sparse Mixture-of-Experts model for functional programming code generation that addresses the complementary failure modes of per-language and multi-language fine-tuning through dedicated language-specific experts and an always-active shared expert. At only 3B active parameters, FPMoE matches models several times its size. Future directions include scaling to additional FP languages, larger expert counts, and reinforcement learning from execution feedback.

\section*{Limitations}

\paragraph{Fixed language set.}
FPMoE's MoE architecture is hardcoded for exactly three functional 
programming languages: Haskell, OCaml, and Scala. Extending the 
model to support additional functional languages (e.g., Erlang, 
Clojure, or F\#) would require retraining the full model rather 
than incrementally adding a new expert, limiting the architecture's 
extensibility.

\paragraph{Router interpretability.}
Although we provide an interpretability analysis of learned expert 
specialization, we do not fully validate that the router dispatches 
tokens strictly according to language identity rather than other 
latent features. A more rigorous probing study—e.g., measuring 
routing entropy conditioned on language—would strengthen the 
claim that specialization emerges by design rather than by 
coincidence.




\paragraph{Training data recency.}
The pre-training corpora (CodeParrot, 2022; AllenGeng, 2024) may 
not reflect the latest library ecosystems or language features, 
particularly for rapidly evolving languages such as Scala~3, which 
introduced substantial syntactic and type system changes. Models 
trained on older corpora may underperform on modern idiomatic 
code.

\paragraph{Memory footprint of MoE inference.}
Although FPMoE activates only 3B or 6B parameters per forward 
pass, the full set of expert weights must reside in memory during 
inference. This increases the memory footprint relative to a 
dense model of equivalent active parameter count, which may 
present deployment challenges in resource-constrained environments.

\newpage

\bibliography{anthology,custom}


\appendix 
\label{sec:appendix}



\section{Hyperparameters for Tuning}
\raggedright
In our tuning pipeline, we used this configuration,
\begin{table}[h]
\centering
\caption{Tuning Hyperparameters}
\resizebox{0.7\linewidth}{!}{ 
\begin{tabular}{ll}
\toprule
\textbf{Hyperparameter} & \textbf{Value} \\
\midrule
torch\_dtype & bfloat16 \\
load\_in\_4bit & true \\
lora\_alpha & 64 \\
lora\_r & 32 \\
lora\_dropout & 0.0 \\
use\_rslora & false \\
learning\_rate & 1e-4 \\
num\_train\_epochs & 1 \\
per\_device\_train\_batch\_size & 32 \\
\bottomrule
\end{tabular}
}
\label{tab:tuning-hyperparams}
\end{table}

\end{document}